\documentclass{emulateapj}
\usepackage{float}

\begin{document}

\title{The Chemical Evolution of Star-Forming Galaxies Over the Last 11 Billion Years}

\author{H. Jabran Zahid$^{1,2}$, Margaret J. Geller$^{2}$, Lisa J. Kewley$^{1,3}$,Ho Seong Hwang$^{2}$, Daniel G. Fabricant$^{2}$ \& Michael J. Kurtz$^{2}$ }
\affil{$^{1}$University of Hawaii at Manoa, Institute for Astronomy - 2680 Woodlawn Dr., Honolulu,  HI 96822, USA}
\affil{$^{2}$Smithsonian Astrophysical Observatory - 60 Garden St., Cambridge, MA 02138}
\affil{$^{3}$Australian National University, Research School of Astronomy and Astrophysics - Cotter Road, Weston Creek, ACT 2611, Australia}

\begin{abstract}
We calculate the stellar mass-metallicity relation at five epochs ranging to $z\sim2.3$. We quantify evolution in the shape of the mass-metallicity relation as a function of redshift; the mass-metallicity relation flattens at late times. There is an empirical upper limit to the gas-phase oxygen abundance in star-forming galaxies that is independent of redshift. From examination of the mass-metallicity relation and its observed scatter we show that the flattening at late times is a consequence of evolution in the stellar mass where galaxies enrich to this empirical upper metallicity limit; there is also evolution in the fraction of galaxies at a fixed stellar mass that enrich to this limit. The stellar mass where metallicities begin to saturate is $\sim0.7$ dex smaller in the local universe than it is at $z\sim0.8$. 

\end{abstract}

\section{Introduction}

Gas dynamics and star formation regulate the gas-phase oxygen abundance (metallicity) of star-forming galaxies. Oxygen in the universe forms in the late stage evolution of massive stars. Oxygen dispersed into the interstellar medium (ISM) by supernovae and stellar winds increases the metallicity of galaxies as they increase their stellar mass. Observations suggest that in most star-forming galaxies at $z\lesssim2$, the increase in stellar mass is dominated by cosmological inflows of gas from the intergalactic medium \citep[among others]{Noeske2007a, Whitaker2012}. At the same time, outflows of gas are ubiquitously observed in star-forming galaxies out to $z\sim3$ \citep{Weiner2009, Chen2010, Steidel2010}. Because metallicity is established by the interplay of gas flows and star-formation, observations of the chemical evolution of galaxies provides important constraints for these physical processes in models of galaxy evolution \citep[e.g.][]{Dave2011b, Zahid2012b, Moller2013}.

The mass-metallicity (MZ) relation is a measure of the average gas-phase oxygen abundance as a function of stellar mass. \citet{Lequeux1979} first observed the relation between stellar mass and metallicity in nearby irregular and blue compact galaxies. From the analysis of $\sim50,000$ star-forming galaxies in the Sloan Digital Sky Survey, \citet{Tremonti2004} establish the MZ relation in the local universe. Subsequent efforts show that the MZ relation extends to low stellar masses \citep[$\sim10^7M_\odot$;][]{Lee2006, Zahid2012a, Berg2012} and out to \emph{at least} $z\sim3$ \citep[among others]{Savaglio2005, Erb2006b, Maiolino2008, Mannucci2009, Zahid2011a, Foster2012, Yabe2012, Perez-Montero2012, Yuan2013}, with indications that it may extend to even higher redshifts \citep{Laskar2011, Moller2013}. The metallicities of galaxies increase with stellar mass and the metallicities at all stellar masses decrease with redshift. Because of systematic uncertainties and a lack of sufficiently large samples of galaxies with well determined metallicities at intermediate and high redshifts, evolution in the shape of the MZ relation is not well known.

A robust determination of the evolution of the MZ relation requires consistent measurements of the metallicities \emph{and} stellar masses of galaxies. Here we bring together a homogeneously analyzed sample of galaxies with well determined MZ relations to investigate the evolution of the MZ relation to $z\sim2.3$. In Section 2 we describe the data and the methods applied to determine metallicities and stellar masses. In Section 3 we present the MZ relation and its scatter and quantify its evolution. We discuss our results in Section 4 and conclude with a summary in Section 5. We assume the standard cosmology $(H_{0}, \Omega_{m}, \Omega_{\Lambda}) = (70$ km s$^{-1}$ Mpc$^{-1}$, 0.3, 0.7) and a \citet{Chabrier2003} IMF.

\section{Data and Methods}

\subsection{The Data}
We derive the  MZ relation at $z=0.08, 0.29$ and 0.78 using data from the Sloan Digital Sky Survey DR7 \citep[SDSS;][]{Abazajian2009}, the Smithsonian Hectospec Lensing Survey \citep[SHELS;][Hwang et al., in prep]{Geller2005} and the Deep Extragalactic Evolutionary Probe 2 Survey DR3 \citep[DEEP2;][]{Davis2003}, respectively. The MZ relation at $z = 1.41$ and 2.26 are determined by \citet{Yabe2012} and \citet{Erb2006b}, respectively. We determine the stellar masses of galaxies in their sample using our method (see Section 2.2) and adopt their metallicity estimates but convert them for consistency with our measurements of metallicity in the $z<1$ samples (see Section 2.3).

Our primary selection criteria for galaxies in the $z<1$ samples are the signal-to-noise (S/N) ratios of strong emission lines. \citet{Foster2012} show that the derived MZ relation does not vary significantly with S/N cuts on most of the strong emission lines used in this study, though S/N cuts on the [OIII]$\lambda5007$ line can lead to a significant bias. The requisite lines for determining metallicity in the $z<1$ samples are [OII]$\lambda3727$, H$\beta$ and [OIII]$\lambda5007$ (see Section 2.3). We require a S/N $>3$ in the  [OII]$\lambda3727$ and H$\beta$ emission lines. Additionally, in the SDSS and SHELS sample we use the H$\alpha$ and [NII]$\lambda6584$ emission lines in selecting star-forming galaxies and apply the same S/N criteria.

The SDSS spectroscopic sample consists of $\sim900,000$ galaxies primarily in the redshift range of $0<z<0.3$. We adopt emission line fluxes measured by the MPA/JHU\footnote{http://www.mpa-garching.mpg.de/SDSS/DR7/} group and determine the stellar masses from the $ugriz$-band photometry. AGN are removed from the sample using the [OIII]/H$\beta$ vs. [NII]/H$\alpha$ diagram \citep[i.e. the BPT method,][]{Baldwin1981, Kauffmann2003, Kewley2006}. We require an aperture covering fraction $\gtrsim20\% $\citep{Kewley2005} and select galaxies with $z<0.12$. Our final sample consists of $\sim51,000$ star-forming galaxies in the redshift range of $0.02<z<0.12$

The SHELS survey consists of $\sim25,000$ galaxies in the F1 (Hwang et al., in prep) and F2 \citep{Geller2005, Hwang2012, Geller2012} regions of the Deep Lens Survey \citep{Wittman2002} spanning the redshift range of $0 < z < 0.8$. We determine stellar masses from the $ugriz$-band SDSS photometry and remove AGN using the BPT method. Our sample consists of 3,577 star-forming galaxies in the redshift range $0.2<z<0.38$. 

The DEEP2\footnote{http://deep.ps.uci.edu/DR3/} survey \citep{Davis2003} consists of $\sim50,000$ galaxies spanning the redshift range of $0.7<z<1.4$. We determine stellar masses from $BRI$-band photometry with additional $K_s$-band photometry for half of the sample. This sample differs slightly from the sample we analyzed in \citet{Zahid2011a}. For consistency with the SDSS and SHELS analysis, we use stellar population synthesis models to remove Balmer absorption. Additionally, we limit AGN contamination by removing 17 x-ray galaxies in the sample \citep{Goulding2012}. Our final sample consists of 1,254 star-forming galaxies in the redshift range $0.75<z<0.82$.

\citet{Yabe2012} and \citet{Erb2006b} measure the MZ relation at $z \sim 1.4$ and $z\sim2.3$, respectively. \citet{Yabe2012} determine the MZ relation from the stacked spectra of 71 objects with significant detections of H$\alpha$ and \citet{Erb2006b} determine the MZ relation from the stacked spectra of 87 UV-selected star-forming galaxies. The MZ relations determined by \citet{Erb2006b} and \citet{Yabe2012} are subject to some systematic uncertainties (see Section 2.3) that we are currently unable to quantify. We include these data to demonstrate that the best data currently available at high redshift are qualitatively consistent with the main conclusions of this study based \emph{solely} on the $z<1$ samples.

\subsection{Stellar Mass Determination}
We measure stellar masses for the five samples using the Le Phare\footnote{$\url{http://www.cfht.hawaii.edu/{}_{\textrm{\symbol{126}}}arnouts/LEPHARE/cfht\_}$ $\url{lephare/lephare.html}$} code developed by Arnouts \& Ilbert. We determine the mass-to-light ratio by fitting the SED with stellar population synthesis models and we scale the luminosity by the mass-to-light ratio to yield a stellar mass estimate \citep[see][]{Bell2003b}. We use the stellar templates of \citet{Bruzual2003} and a \citet{Chabrier2003} IMF. The models have two metallicities and seven exponentially decreasing star formation models (SFR $\propto e^{-t/\tau}$) with $\tau = 0.1,0.3,1,2,3,5,10,15$ and $30$ Gyrs. We apply the extinction law of \citet{Calzetti2000} allowing E(B$-$V) from 0 to 0.6 and a stellar population age range from 0 to 13 Gyrs. For each of our samples, the age never exceeds the age of the universe at the median redshift of the sample. 

We adopt the median of the mass distribution as our estimate of the stellar mass. We compare the stellar masses we determine for our SDSS sample with those derived by the MPA/JHU group. We find a 0.17 dex dispersion after correcting for a systematic offset between the two methods. This dispersion is consistent with the observational uncertainty in the photometry.

\subsection{Metallicity Determination}

\begin{figure*}
\begin{center}
\includegraphics[width=2\columnwidth]{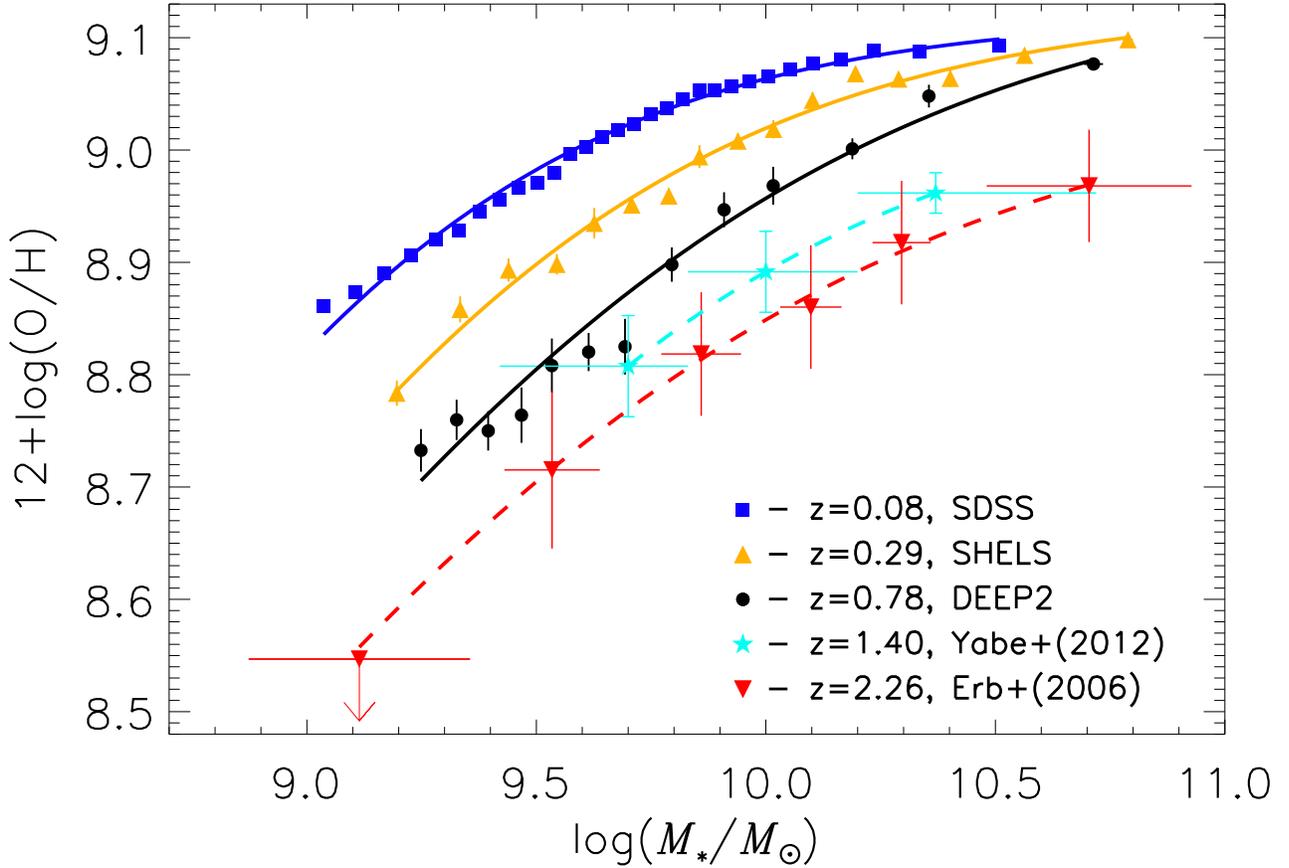}
\end{center}
\caption{The MZ relation at five epochs ranging to $z\sim2.3$. The curves are fits to the data defined by Equation \ref{eq:fit}. The solid curves indicate metallicities determined using the KK04 strong-line method and the dashed curves indicate metallicities converted using the formulae of \citet{Kewley2008}. \emph{Data presented in this figure can be obtained from HJZ upon request.}}
\label{fig:mzr}
\end{figure*}

Ratios of the strength of collisionally excited emission lines to recombination lines are both theoretically and empirically calibrated for determining metallicity. A long-standing astrophysical problem is that the absolute metallicity determined from theoretical calibrations is typically $\sim0.3$ dex higher than metallicities determined using empirical calibration methods. However, we emphasize that our analysis requires a robust \emph{relative} estimate of the metallicity, which the various metallicity diagnostics deliver \citep{Kewley2008}. \citet{Kewley2008} provide a table of formulae allowing for conversion of metallicities into various baseline methods. 

We determine metallicities from the $R23$ line ratio calibrated by \citet[hereafter KK04]{Kobulnicky2004}. A major advantage of this method is that it explicitly solves and corrects for the ionization parameter. The relevant ratios of measured emission line intensities are
\begin{equation}
R23 = \frac{\mathrm{[OII]}\lambda3727 + \mathrm{[OIII]}\lambda4959, 5007}{\mathrm{H}\beta}
\end{equation}
and 
\begin{equation}
O32 = \frac{\mathrm{[OIII]}\lambda4959, 5007}{\mathrm{[OII]}\lambda3727}.
\end{equation}
We assume that the ratio of the fluxes of [OIII]$\lambda5007$ to [OIII]$\lambda4959$ is 3 \citep{Osterbrock1989} and adopt a value of 1.33 times the [OIII]$\lambda5007$ when summing the [OIII]$\lambda5007$ and [OIII]$\lambda4959$ line strengths. The $R23$ method is sensitive to the ionization state of the gas and the $O32$ ratio is used to correct for variations. We apply this method to the SDSS, SHELS and DEEP2 data. The galaxies in this study all lie on the upper metallicity branch \citep[see][]{Zahid2011a}. 

At higher redshifts the requisite lines for determination of the $R23$ ratio are rarely observed. \citet[hereafter PP04]{Pettini2004} calibrate 
\begin{equation}
N2 = (\mathrm{[NII]}\lambda6584/\mathrm{H}\alpha)
\end{equation}
to yield the metallicity. The advantage of this diagnostics is that the lines are closely spaced and easily observed in a single spectroscopic setting. However, diagnostics using [NII]$\lambda6584$ are known to depend on the ionization parameter \citep{Kewley2002} and the N/O ratio which is not constant with metallicity \citep[e.g.][]{Perez-Montero2009b}. These effects may evolve with redshift (Kewley et al., in  prep). Therefore, an MZ relation based on this diagnostics may be systematically biased. Both \citet{Erb2006b} and \citet{Yabe2012} determine metallicities using the PP04 $N2$ calibration. We convert metallicities determined from the PP04 $N2$ diagnostic to the KK04 diagnostic using the conversion formulae given in \citet{Kewley2008}.

\section{Results}

\subsection{The MZ Relation}

\begin{deluxetable*}{cccccc}
\tablewidth{410pt}
\tablecaption{MZ Relation Fit}
\tablehead{\colhead{Sample} & \colhead{Redshift} &\colhead{$Z_o$} & \colhead{$\mathrm{log}(M_o/M_\odot)$} & \colhead{$\gamma$} & \colhead{Calibration} }
\startdata
SDSS     & 0.08 & 9.121 $\pm$ 0.002 & 8.999 $\pm$ 0.005 & 0.85 $\pm$ 0.02 & KK04 \\
SHELS   & 0.29 & 9.130 $\pm$ 0.007 & 9.304 $\pm$ 0.019 & 0.77 $\pm$ 0.05 & KK04 \\
DEEP2   & 0.78 & 9.161 $\pm$ 0.026 & 9.661 $\pm$ 0.086 & 0.65 $\pm$ 0.07 & KK04 \\
Y12        & 1.40 & 9.06 $\pm$ 0.36    & 9.6  $\pm$ 0.8   & 0.7  $\pm$ 1.5 & PP04 \\
E06        & 2.26 & 9.06 $\pm$ 0.27    & 9.7   $\pm$ 0.9   & 0.6 $\pm$ 0.7  & PP04 \\

\enddata
\label{tab:data}
\tablecomments{The sample and median redshift are given in columns 1 and 2, respectively. The fit parameters from Equation \ref{eq:fit} are given in columns 3-5. Column 6 indicates the strong line method used for deriving metallicity. We convert PP04 metallicities to the KK04 calibration using the formulae from \citet{Kewley2008}.}
\end{deluxetable*}

Figure \ref{fig:mzr} shows the MZ relations at five epochs. We determine the MZ relation for the SDSS, SHELS and DEEP2 samples by binning the data. We sort galaxies into equally populated bins of stellar mass and plot the median stellar mass and metallicity for each bin. The MZ relation of \citet{Yabe2012} and \citet{Erb2006b} is determined from stacked spectra sorted by stellar mass. The errors for the $z<1$ data are determined from bootstrapping. For the $z>1$ data the errors are determined from the dispersion in the stacked spectra. 

We fit the MZ relation using the function defined by \citet{Moustakas2011}. The functional form of the MZ relation fit is
\begin{equation}
\mathrm{12 + log(O/H)} = Z_o - \mathrm{log}\left[1 + \left(\frac{M_\ast}{M_o}\right)^{- \gamma} \right].
\label{eq:fit}
\end{equation}
This function is desirable because it is monotonic unlike the commonly used quadratic fit \citep[e.g.][]{Tremonti2004, Zahid2011a} which turns over at high stellar masses. Furthermore, the parameters of the fit reflect our physical intuition of chemical evolution and are more straightforward to interpret physically \citep[see discussion in appendix of][]{Moustakas2011}. In Equation \ref{eq:fit}, $Z_o$ is the asymptotic metallicity where the MZ relation flattens, $M_o$ is the characteristic mass where the MZ relation begins to flatten and $\gamma$ is the power law slope of the MZ relation for $M_\ast << M_o$. The fitted value of $Z_o$ is subject to uncertainties in the absolute calibration of the metallicity diagnostic, though the relative values are robust (see Section 2.3). We do not probe stellar masses where $M_\ast << M_o$. Therefore the power law slope of the MZ relation at the low mass end, $\gamma$, is not well constrained. Table 1 lists the fitted parameters. We propagate the observational uncertainties to the parameter errors.

\subsection{Scatter in the MZ Relation}

\begin{figure*}
\begin{center}
\includegraphics[width=2\columnwidth]{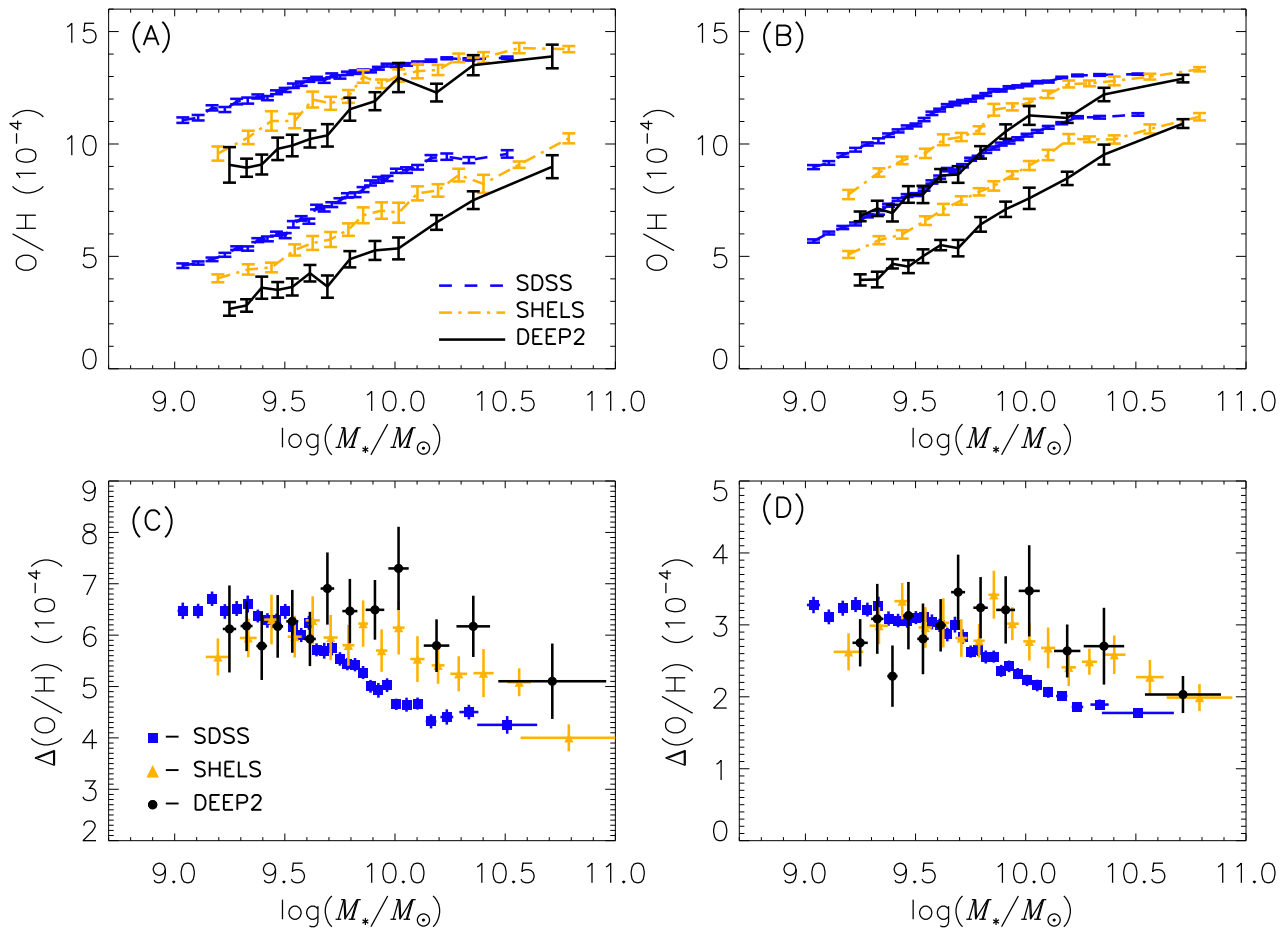}
\end{center}
\caption{Limits containing the central (A) 85\% and (B) 50\% of the metallicity distribution and the difference between the upper and lower limits for the central (C) 85\% and (D) 50\% of the galaxy metallicity distribution as a function of stellar mass for the $z<1$ samples. For clarity the metallicity is plotted on a linear scale.}
\label{fig:scatter}
\end{figure*}

The scatter in the metallicity distribution as a function of stellar mass and redshift provides important additional constraints for the chemical evolution of galaxies. In Figure \ref{fig:scatter} we plot the scatter in the MZ relation. We note that we have converted the metallicity to linear units for clarity. In Figures \ref{fig:scatter}A and \ref{fig:scatter}B we plot the limits containing the central 85\% and 50\% of the galaxy metallicity distribution, respectively, as a function of stellar mass. The errors bars are determined from bootstrapping the sample distribution. In Figures \ref{fig:scatter}C and \ref{fig:scatter}D we plot the scatter in the MZ relation (defined as the difference between the upper and lower limits of the 85\% and 50\% contour, respectively) as a function of stellar mass. The error bars are the bootstrapped errors on the limits added in quadrature.

\section{Discussion}

\subsection{Evolution of the MZ Relation}

Figure \ref{fig:mzr} shows that at a fixed stellar mass (over the stellar mass range $10^{9}M_\odot \lesssim M_\ast \lesssim 10^{11}M_\odot$) the average metallicity of the star-forming galaxy population increases as a function of cosmic time since $z\sim2.3$. However, the evolution is mass dependent; thus the shape of the MZ relation evolves with time. There is a greater metallicity evolution in lower mass galaxies since $z\sim0.8$ as compared to more massive galaxies. The smaller difference in the MZ relation for massive galaxies since $z\sim0.8$ is due to flattening of the MZ relation at higher stellar masses.  

Figures \ref{fig:scatter}A and \ref{fig:scatter}B show that the flattening in the MZ relation at late times results from a mass dependent flattening of the upper envelope of the metallicity distribution. The stellar mass where galaxy metallicities saturate decreases at late times. The statistically significant evolution of $M_o$ to lower stellar masses at late times (see Table 1) quantifies this effect. The metallicity distribution in Figure \ref{fig:scatter}A and \ref{fig:scatter}B also shows that there is an empirical upper metallicity limit. $Z_o$ in Equation \ref{eq:fit} quantifies the upper metallicity limit. Within the observational uncertainties, $Z_o$ is constant to $z\sim2.3$. 

Analytical models provide a physical interpretation of $Z_o$ and the observed effect of metallicity saturation. In a closed-box model of chemical evolution where no gas enters or leaves the system, the gas-phase oxygen abundance can evolve to arbitrarily high values. However, galaxies do not evolve as closed boxes. More realistically, in an inflow model of chemical evolution where star formation is fueled by inflowing, pristine gas, the gas-phase oxygen abundance saturates at a metallicity equivalent to the nucleosynthetic stellar yield\footnote{For reference, the saturation metallicity (i.e. gas-phase oxygen abundance) in Figure \ref{fig:scatter} converted to mass density units is $\sim0.013$. Typical theoretical values of the nucleosynthetic stellar yield of oxygen in the same units vary between 0.01 and 0.02.} \citep{Edmunds1990}. In this case, the amount of oxygen produced by massive stars is balanced by the amount of oxygen locked-up forever in low mass stars. When outflows are included in analytical models, the upper metallicity limit can \emph{potentially} be reduced to an effective stellar yield \citep[e.g.][]{Edmunds1990}. This yield depends on the composition of the outflow, which is not yet well constrained by observations. The constancy of $Z_o$ suggests that the metallicity of outflows, at a fixed stellar mass, has not evolved significantly since $z\sim2.3$.

Figures \ref{fig:scatter}A and \ref{fig:scatter}B show that the fraction of saturated galaxies at a fixed stellar mass increases at late times. Thus massive galaxies in Figures \ref{fig:scatter}C and \ref{fig:scatter}D have smaller scatter in their metallicity distribution at late times. The straightforward interpretation of the data is that at early times ($z\sim0.8$) the scatter in star-forming galaxy metallicities is nearly constant as a function of stellar mass. As galaxies evolve, metallicity saturation occurs first in the most massive galaxies and later at lower stellar masses. The stellar mass where the metallicity saturates and the fraction of saturated galaxies at a fixed stellar mass evolves with redshift. The data show that metallicity saturation leads to a decrease in the scatter with time (Figures \ref{fig:scatter}C and \ref{fig:scatter}D). Therefore, the observed scatter in the MZ relation at $z\sim0.8$ is likely to be a lower limit to the scatter in the MZ relation at higher redshifts.

\subsection{Comparison with Previous Observational Studies}

Evolution in the shape of the MZ relation has previously been reported \citep{Savaglio2005, Maiolino2008, Zahid2011a}. At a fixed stellar mass, there is relatively greater enrichment in less massive galaxies since $z\sim0.8$. Our results provide a robust quantification of this evolution. 

Galaxies are observed to form in ``downsizing" such that massive galaxies form earlier and more rapidly than low mass systems and the star-forming population of galaxies is dominated by less massive galaxies at later times \citep{Cowie1996}. Flattening of the slope of the MZ relation at late times has been interpreted as the chemical version of galaxy downsizing \citep{Savaglio2005, Maier2006, Maiolino2008, Zahid2011a}. We show that flattening of the MZ relation is due to a redshift evolution in the stellar mass where the metallicities of galaxies begin to saturate. This is probably due to gradual depletion of gas reservoirs at late times for galaxies at a fixed stellar mass. In this case, the flattening of the slope of the MZ relation can be interpreted without the need to invoke downsizing.

\citet{Moustakas2011} study the MZ relation to $z\sim0.75$ using a sample of $\sim3,000$ galaxies drawn from the AGN and Galaxy Evolution Survey \citep[AGES,][]{Kochanek2012}. In contrast to our results, they find that the shape of the MZ relation does not evolve in a mass-dependent way for galaxies with stellar masses $>10^{9.8}M_\odot$. They base their conclusion on an MZ relation which is only derived for the most massive galaxies at intermediate redshifts. At $z\sim0.6$, the lowest stellar mass bin has a median stellar mass of $10^{10.6} M_\odot$ as compared to $10^{9.2} M_\odot$ for the DEEP2 galaxy sample at $z\sim0.8$. Thus the shape of the MZ relation at intermediate redshifts is not well constrained by the data. However, a direct comparison of the data suggests that systematic differences in sample selection may be the source of the discrepancy. The MZ relation measured by \citet{Moustakas2011} at $z\sim0.3$ is significantly flatter for galaxies with $M_\ast>10^{10.2}M_\odot$ as compared to the relation we derive from the SHELS sample at the same redshift. \citet{Moustakas2011} select galaxies with well measured [OIII]$\lambda5007$. \citet{Foster2012} show that a S/N cut on the [OIII]$\lambda5007$ emission line leads to a flattening of the MZ relation for massive galaxies because metal-rich galaxies have weak oxygen emission and low ionization parameter \citep{Zahid2012a}. A systematic bias against massive, metal-rich galaxies could lead to an underestimate of the relation for massive galaxies and may explain the lack of evolution in the shape reported by \citet{Moustakas2011}.

\section{Summary and Conclusions}

We investigate the evolution of the mass-metallicity relation and its shape using five samples spanning the redshift range of $0 \lesssim z \lesssim2.3$. We calculate the stellar masses and metallicities for all five samples using as consistent a methodology as is currently possible. Our conclusions are primarily based on the three $z<1$ galaxy samples at $z = 0.08, 0.29$ and $0.78$ for which we have metallicities measured in individual galaxies. We show that the MZ relation at $z = 1.4$ and $2.26$ determined from stacked spectra are consistent with the main conclusions of this study which are:

\begin{itemize}
\item{The metallicities of star-forming galaxies at a fixed stellar mass decrease at all stellar mass $\gtrsim10^{9}M_\odot$ as a function redshift.}

\item{The MZ relation since $z\sim0.8$ evolves in a mass dependent manner such that the shape of the MZ relation changes with redshift.}

\item{Galaxy metallicities saturate. The stellar mass where galaxy metallicities saturate and the fraction of galaxies with saturated metallicities at a fixed stellar mass evolves. Thus there is a mass dependent decrease in the scatter and a flattening of the MZ relation at late times.}

\item{We attribute the flattening of the MZ relation at late times to an empirical upper limit in the gas-phase oxygen abundance for star-forming galaxies which does not evolve significantly.}
\end{itemize}

We quantify evolution in the shape of the MZ relation out to $z\sim0.8$. The multiplexing capability of the new generation of near-infrared spectrographs will allow us to rigorously establish the MZ relation for $z>1$ galaxies. 

\acknowledgements
We thank the referee for useful comments for improving the paper, John Moustakas, Vivienne Wilde and Nelson Caldwell for providing useful advice and routines and Kiyoto Yabe and Dawn Erb for sharing their data. We thank Susan Tokarz, Sean Moran and Warren Brown for the careful reduction of the SHELS data and Mike Calkins and Perry Berlind for operating Hectospec. HJZ is grateful to Freeha Riaz and Ananda Wickramsekera for their warm and generous hospitality. HJZ and LJK acknowledge support by NSF EARLY CAREER AWARD AST07-48559. 

{\it Facilities:} \facility {MMT(Hectospec)} 


 \end{document}